# Database Reverse Engineering based on Association Rule Mining

**Nattapon Pannurat[1], Nittaya Kerdprasop[2] and Kittisak Kerdprasop[2],\***

[1] **Faculty of Information Sciences, Nakhon Ratchasima College**
290 Moo 2, Mitraphap Road, Nakhon Ratchasima, 30000, Thailand

[2] **Data Engineering and Knowledge Discovery Research Unit, Suranaree University of Technology**
111 University Avenue, Nakhon Ratchasima, 30000, Thailand

**Abstract**
Maintaining a legacy database is a difficult task especially when system documentation is poor written or even missing. Database reverse engineering is an attempt to recover high-level conceptual design from the existing database instances. In this paper, we propose a technique to discover conceptual schema using the association mining technique. The discovered schema corresponds to the normalization at the third normal form, which is a common practice in many business organizations. Our algorithm also includes the rule filtering heuristic to solve the problem of exponential growth of discovered rules inherited with the association mining technique.
***Keywords:*** *Legacy Databases, Reverse Engineering, Database Design, Database Normalization, Association Mining.*

## 1. Introduction

Legacy databases are obviously valuable assets to many organizations. These databases were mostly developed with technologies in the 1970s [14] using old programming languages such as COBOL and RPG, and file systems of the mini-computer platforms. Some databases were even designed with the outdated concepts such as hierarchical data model, and thus made them difficult to be maintained and adjusted to serve current needs of modern companies.

One solution to modernize the legacy databases is to migrate and transform their structures and corresponding contents to the new systems. This approach is, however, hard to achieve if the design document of the system does no longer exist, which is the common situation in most enterprises. To solve the problems of recovering database structures and migrating legacy databases, we propose the database reverse engineering methodology.

The process of reverse engineering [7] originally aimed at discovering design and production procedure from devices, end products, or other hardware. This methodology often used in the Second World War for military advantage by copying opponents' technologies. Reverse engineering of software refers to the process of discovering source code and system design from the available software [7].

In database community, reverse engineering is an attempt to extract the domain semantics such as keys, functional dependencies and integrity constraints from the existing database structures [6, 13]. Typically, database reverse engineering is the process of extracting design specifications from legacy systems and making the reverse transformation from logical to conceptual schema [6, 15].

Our work deals with the reverse schema process by making a step further from logical schema to the lower level of database instances. We apply the machine learning technique, association rule mining in particular, to induce dependency relationships among data attributes. The major problem of applying association mining to real-life databases is that it always generates tremendous amount of association rules [11, 12]. We thus include the rule- filtering component in our design to select only promising association rules.

The structure of this paper is organized as follows. Section 2 presents the basic concept and the design framework of our methodology. Section 3 explains the system implementation by means of an example. Section 4

---

\* Corresponding author

discusses related work. Finally, Section 5 concludes the paper.

## 2. Database Reverse Engineering with NoWARs

The objective of our system is to induce conceptual schema from the database instances with the basic assumption that database design documents are absent. We apply the normalization principles and the association mining technique to discover the missing database design.

Normalization [8] is the process to transform unstructured relation into separate relations, called normalized ones. The main purpose of this separation is to eliminate redundant data and reduce data anomaly (insert, update, and delete). There are many different levels of normalization depending on the purpose of database designer. Most database applications are designed to be in the third and the Boyce-Codd normal forms in which their dependency relations [3] are sufficient for most organizational requirements. Figure 1 [9] illustrates the refinement steps from un-normalized relations to the relations in fifth normal form.

The main condition to transform from one normal form to the next level is the dependency relationship, which is a constraint between two sets of attributes in a relation. Experienced database designers are able to elicit this kind of information. But in the reverse engineering process in which the business process and operational requirements are unknown, this task of dependency analysis is tough even for the experienced ones. We thus propose to use the machine learning technique called association mining.

Association mining searches for interesting relationships among a large set of data items [1, 2]. The discovery of interesting association relationships among huge amounts of business transaction records can help in many business decision making process, such as catalog design, cross-marketing, and loose-leader analysis [11]. An example of association rule mining is market basket analysis. This process analyzes customer buying habits by finding association the different items that customers place in their shopping baskets. For example, customers are buying milk also tend to buy bread and water drinking at the same time. These can represent in association rule as follows.

milk [support=5%] → bread, water drinking [support=5%]
[confidence=100%]

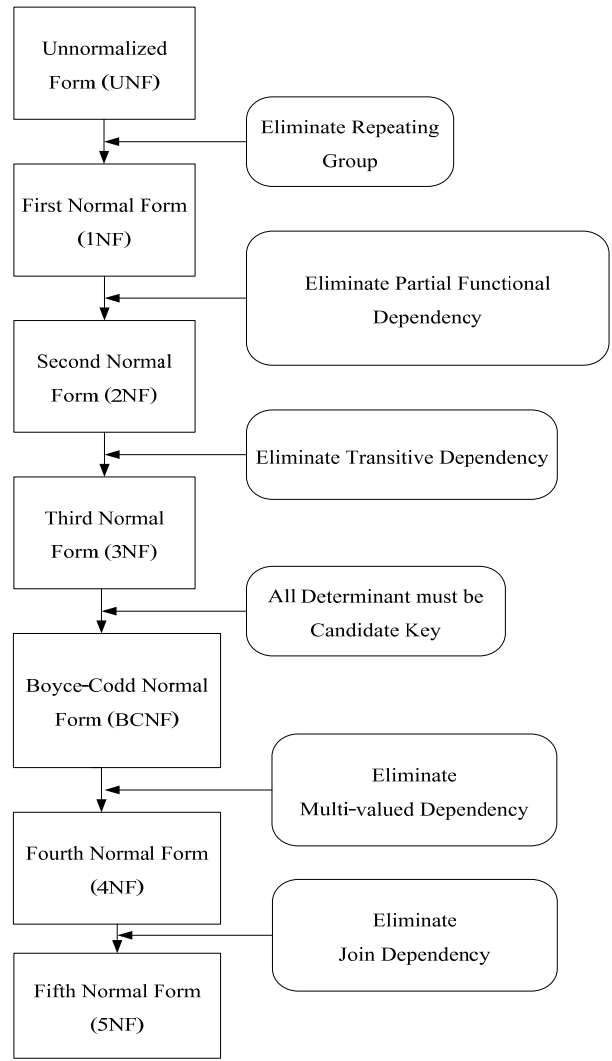

Fig. 1 Normalization steps.

A support of 5% for association rule means that 5% of all the transactions under analysis show that milk, bread, and water drinking are purchased together. A confidence of 100% means that 100% of the customers who purchased some milk also bought bread and water drinking.

Our methodology of database reverse engineering composes of designing and improving the process of normalization with association analysis technique. We use the normalization concept and association analysis technique to create a new algorithm called NoWARs (Normalization With Association Rules).

NoWARs is an algorithm that combines normalization process and association mining technique together. We can find association rules by taking the dataset on a database and feeding into a data mining process. We use

Apriori algorithm to find association rules. NoWARs has two important steps, first finding association rules and second normalization with rules obtained from the first step. The details of NoWARs algorithm are shown in Figure 2 and its workflow are shown in Figure 3.

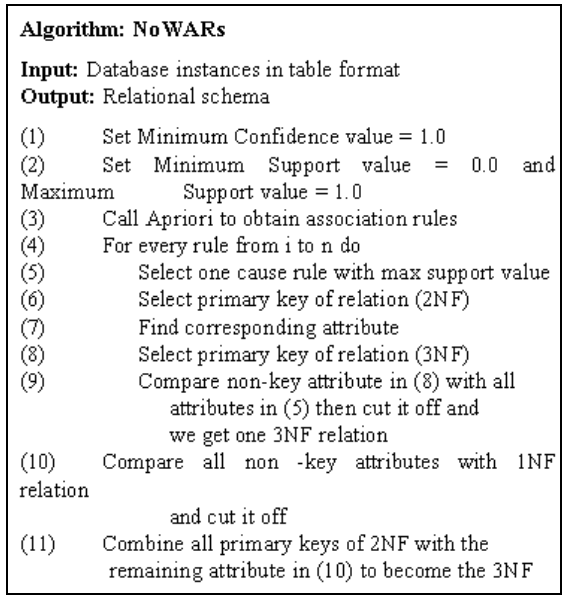

Fig. 2 NoWARs algorithm.

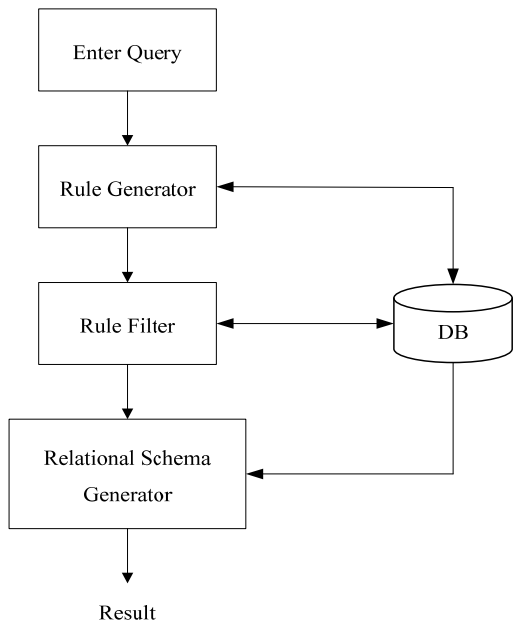

Fig. 3 The work flow of NoWARs algorihm.

## 3. Implementation

The algorithm NoWARs starts when user enter query to define dataset to normalize. Then NoWARs will find the association rules by calling Apriori algorithm and save resulting a form of association rules in the database. Then NoWARs will select some rules to use in normalization process. Finally, use the selected rules to generate the 3NF table in relational schema form. The input of NoWARs is the un-normalize table. The example of input data format is shown in Table 1.

Table 1: Example of input data.

| INV | DATE | C_ID | P_ID | P_Name | QTY |
|---|---|---|---|---|---|
| 001 | 9/1/2010 | C01 | P01 | Printer | 3 |
| 001 | 9/1/2010 | C01 | P02 | Phone | 5 |
| 002 | 9/1/2010 | C03 | P05 | TV | 6 |
| 002 | 9/1/2010 | C03 | P04 | Lamp | 2 |
| : | : | : | : | : | : |

The un-normalized data as shown in Table 1 will be analyzed by the algorithm, and then its schema in 3NF is generated. We perform experimentation with five datasets as shown in Table 2. We use Oracle Database 10g XE Edition, tested on Pentium IV 3.0 GHz with RAM 512 MB machine.

Table 2: Number of records and attributes in experimental datasets.

| Dataset Name | Number of Records | Number of attributes |
|---|---|---|
| Register | 12438 | 157 |
| Video_Rental | 483478 | 523 |
| Data_Org | 91845 | 334 |
| Invoice | 119795 | 123 |
| Car_Color | 199337 | 312 |

We take the Register dataset as a running example. This dataset is originally un-normalized and its structure is as follows.

Register (STUDENT_CODE, STUDENT_NAME,
 TEACHER_CODE, TEACHER_NAME,
 UNIT, SUBJECT_CODE, SUBJECT_NAME)

After execution, its conceptual schema is recovered as shown in Figure 4. The performance of rule-filtering also analyzed and shown in Figure 5.

IJCSI International Journal of Computer Science Issues, Vol. 7, Issue 2, No 3, March 2010
www.IJCSI.org
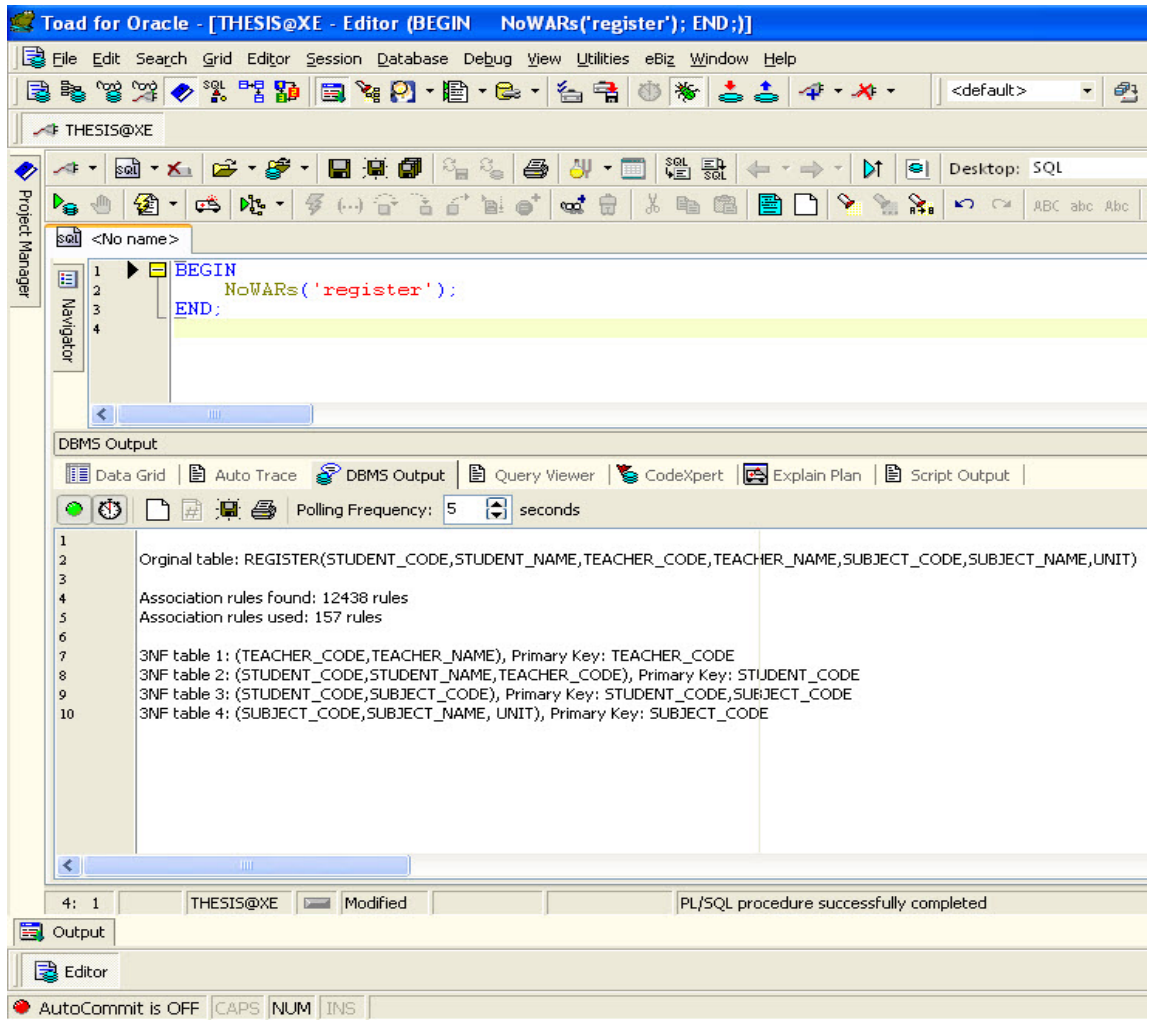

Fig. 4  The result of running NoWARs algorihm on Register dataset

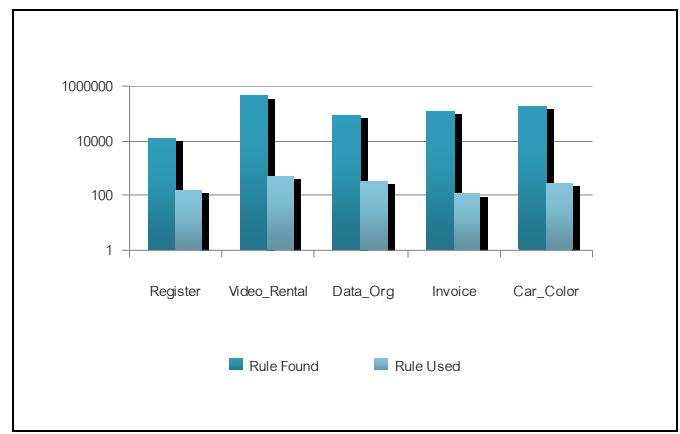

Fig. 5  Performance of rule-filtering component of NoWARs algorithm in reducing number of association rules

## 4. Related Work

Since the introduction of a famous association technique known as Apriori algorithm [1, 2], there have long been immense attempts to integrate this technique to improve database design, consistency checking, and querying. Han et al. [10] improved the DBMiner system to work with relational databases and data warehouses. DBMiner can do many data mining tasks such as classification, prediction and association. Sreenath, Bodagala, Alsabti, and Ranka [16] adopted Apriori algorithm to work with relational database system. They created Fast UPdate algorithm to search association data when the system has new transaction. Tsechansky, Pliskin, Rabinowitz and Porath [17] applied Apriori to find association data from many relations in the database. Berzal, Cubero, Marín and Serrano [4] used Tree-Based Association Rule mining (TBAR) to find association data in relational database. They kept large item set in tree structure format to reduce time cost in association process. Hipp, Güntzer and Grimmer [12] implemented Apriori algorithm with C++ programming language to work on DB2 database system. They used the program to find association data in Daimler-Chrysler Company database.

In parallel to the attempts of applying learning techniques to existing large databases, researchers in the area of database reverse engineering have proposed some means of extracting conceptual schema. Lee and Yoo [14] proposed a method to derive a conceptual model from object-oriented databases. The derivation process is based on forms including business forms and forms for database interaction in the user interface. The final products of their method are the object model and the scenario diagram describing a sequence of operations. The work of Perez et al. [15] emphasized on relational object-oriented conceptual schema extraction. Their reverse engineering technique is based on a formal method of term rewriting. They use terms to represent relational and object-oriented schemas. Term rewriting rules are then generated to represent the correspondences between relational and object-oriented elements. Output of the system is the source code to migrate legacy database to the new system.

Recent work in database reverse engineering has not concentrated on a broad objective of system migration. Researchers rather focus their study on a particular issue of semantic understanding. Lammari et al. [13] proposed a reverse engineering method to discover inter-relational constraints and inheritances embedded in a relational database. Chen et al. [5] also based their study on entity-relationship model. They proposed to apply association rule mining to discover new concepts leading to a proper design of relational database schema. They employed the concept of fuzziness to deal with uncertainty inherited with the association mining process. Our work is also in the line of association mining technique application to the database design. But our main purpose is for the understanding of legacy databases and our method deals with uncertainty by means of heuristic in the step of rule filtering.

## 5. Conclusions and Future Work

A forward engineering approach to the design of a complete database starts from the high-level conceptual design to capture detail requirements of the enterprise. Common tool normally used to represent these requirements is the entity-relationship, or ER, diagram and the product of this design phase is a conceptual schema. Typically, the schema at this level needs some adjustments based on the procedure known as normalization in order to reach a proper database design. Then, the database implementation moves to the lower abstraction level of logical design in which logical schema is constructed in a form of relations, or database tables.

In legacy systems that design documents are incomplete or even missing, the system maintenance or modification is a difficult task due to the lack of knowledge regarding high-level design of the system. To tackle this problem, a database reverse engineering approach is essential.

In this paper, we propose a method to discover conceptual schema from the database instances, or relations. The discovering technique is based on the association mining incorporated with some heuristic to produce a minimal set of association rules. Transformation rules are then applied to convert association rules to database dependencies. Normalization is the principal concept of our heuristic and transformation. To deduce repeating group, insert anomaly, delete anomaly and update anomaly. We introduce the novel algorithm, called NoWARs, to normalize the database tables. In the normalization process, NoWARs uses only 100% confidence association rules with any support values. The results from the NoWARs algorithm are the same as the design schema obtained from the database designer. But NoWARs cannot normalize data model to the level higher than third normal form, which might be the desired level of a highly secured database. We thus plan to improve our methodology to discover a conceptual schema up to the level of fifth normal form.

### Acknowledgments

This research has been conducted at the Data Engineering and Knowledge Discovery (DEKD) research unit, fully

supported by Suranaree University of Technology. The work of first and second authors has been funded by grants from Suranaree University of Technology and the National Research Council of Thailand (NRCT), respectively. The third author has been supported by a grant from the Thailand Research Fund (TRF, grant number RMU5080026).

**Natthapon Pannurat** received his bachelor and master degrees in computer engineering in 2007 and 2009, respectively, from Suranaree University of Technology. He is currently a faculty member of Information Sciences, Nakhon Ratchasima College. His research interests are database management systems, data mining and machine learning.

**Nittaya Kerdprasop** is an associate professor at the school of computer engineering, Suranaree University of Technology, Thailand. She received her B.S. from Mahidol University, Thailand, in 1985, M.S. in computer science from the Prince of Songkla University, Thailand, in 1991 and Ph.D. in computer science from Nova Southeastern University, USA, in 1999. She is a member of ACM and IEEE Computer Society. Her research of interest includes Knowledge Discovery in Databases, Artificial Intelligence, Logic and Constraint Programming, Deductive and Active Databases.

**Kittisak Kerdprasop** is an associate professor and the director of DEKD research unit at the school of computer engineering, Suranaree University of Technology, Thailand. He received his bachelor degree in Mathematics from Srinakarinwirot University, Thailand, in 1986, master degree in computer science from the Prince of Songkla University, Thailand, in 1991 and doctoral degree in computer science from Nova Southeastern University, USA, in 1999. His current research includes Data mining, Machine Learning, Artificial Intelligence, Logic and Functional Programming, Probabilistic Databases and Knowledge Bases.